\documentclass[11pt,a4paper]{article}
\usepackage{t1enc}
\usepackage[latin1]{inputenc}
\usepackage[english]{babel}
\usepackage{dcpic,pictex}
\usepackage{harvard}
\usepackage{amssymb,stmaryrd,mathabx}

\newcommand{\C}{\mathbb C}
\newcommand{\R}{\mathbb R}

\newcommand{\M}{\mathbb M}

\newcommand{\beq}{\begin{equation}}
\newcommand{\eeq}{\end{equation}}
\newcommand{\beqarr}{\begin{eqnarray}}
\newcommand{\eeqarr}{\end{eqnarray}}
\newcommand{\beqa}{\begin{eqnarray*}}
\newcommand{\eeqa}{\end{eqnarray*}}
\begin{document}

\title{\bf Weyl's search for a difference between `physical' and `mathematical' automorphisms }
\author{Erhard Scholz}
\date{July 11, 2016 } 
\maketitle
\thispagestyle{empty}
\begin{abstract}
During his whole scientific life Hermann Weyl  was fascinated by the interrelation  of  physical and mathematical theories. From the mid 1920s onward he  reflected  also on the typical difference between the two epistemic fields and tried to identify  it by comparing  their respective automorphism structures. In a talk given at the end of the 1940s (ETH, Hs 91a:31) he gave the most detailed and coherent discussion of his thoughts on this topic. This paper presents his arguments in the talk and puts it in the context of the later development of gauge theories.
\end{abstract}

{\small \tableofcontents}

\newpage
\section*{Introduction}
\addcontentsline{toc}{section}{\protect\numberline{}Introduction}
The English translation of  {\em Philosophy of Mathematics and Natural Science} \cite{Weyl:PMNEnglish}, in the following quoted as PMN, gave Weyl the opportunity   to reflect once more on the relation between  mathematical and physical automorphisms (often called symmetries) and its meaning for understanding the basic features of nature. This topic had occupied Weyl during his whole scientific life. In a talk {\em Similarity and congruence: a chapter in the epistemology of science} preserved in his Z\"urich Nachlass
 \cite{Weyl:Hs91a:31} we find a coherent balance  of Weyl's thoughts on this theme at the time 1948/49.\footnote{The lecture is undated, but during the talk  Weyl mentioned that his proposal for a unified field theory was made ``just 30 years ago''.} Some parts of this manuscript are identical with passages in PMN, others (but less) are close to formulations which Weyl later used in his book on {\em Symmetry} \cite{Weyl:Symmetry}. All in all we have here a kind of bridge text between his two famous books. Many aspects discussed in the lecture are also present in PMN at different places and therefore not necessarily new for the informed reader, but the concentration on the topic of mathematical versus physical automorphisms gives the manuscript a uniqueness and coherence which justifies a separate discussion of its content. With regard to  {\em Symmetry}  the manuscript can be read as an epistemologically and physically  deeper reflection of the same considerations which lay  at the base of \cite{Weyl:Symmetry}. The text of the lecture is going to  be published in the forthcoming  third edition of the German translation  \cite{Weyl:Symmetrie}.

The lecture concentrates on a topic which was of great interest to Weyl: the question of how to distinguish between mathematical and physical automorphisms in 20th century  physics. It is mentioned in \cite{Weyl:Symmetry} and more extensively discussed in PMN.\footnote{\cite[72f., 82--84]{Weyl:PMNEnglish}} In the talk it is presented as a whole and discussed with even more details than  in PMN.  Weyl argued more clearly for the necessity to distinguish between the  automorphism structures of mathematical and physical theories. He  accentuated the difference by even speaking of the ``group of  automorphisms  of the physical world''. Although this sounded like  an ontological claim, Weyl's arguments  basically pursued an epistemological interest (indicated already in the title of the paper).

This contribution presents Weyl's arguments in favour of taking automorphisms seriously for characterizing objectivity (section \ref{section objectivity}), his distinction between mathematical and physical automorphisms of classical geometry  and physics (section  \ref{section classical}) and the shifts  arising from general  relativity (GR) and ``early'' (pre 1950) quantum theory (sections \ref{section shock}, \ref{section gauge group}). 
Section \ref{section Weyls distinction} proposes an interpretation  of how we may express Weyl's distinction in more general terms; section 6 makes a first step into  discussing later developments from this perspective. 
At the end of his talk Weyl reflected the special nature of the Lorentz group 
(generalizing the ``rotations'' of classical geometry) which had survived the transition from classical to relativistic physics (section  \ref{section Lorentz group}).
A short outlook on later developments, not yet to be seen by Weyl at the end of the 1940s, follows and relates Weyl's arguments to more recent discussion on symmetries in physics and its  philosophy (section  \ref{section discussion}).

\section{\small Automorphisms and objectivity  \label{section objectivity}}
The talk shows how Weyl intended to demarcate  the distinction 
 between mathematical and physical automorphism of physical theories. In  classical  geometry and physics, physical automorphisms  could be based on  the material operations used for defining the  elementary equivalence concept of congruence (``equality and similitude'' in the old terminology).
But Weyl started even more generally, with Leibniz' explanation of the {\em similarity} of two objects, ``two things are similar if they are indiscernible when each is considered by itself (Hs, p. 1)'',\footnote{A similar passage is also to be found in \cite[127f.]{Weyl:Symmetry}. Page quotes referring to the manuscript  \cite{Weyl:Hs91a:31} are abbreviated by (Hs, p. xx). }
and remarks on the Clarke-Leibniz correspondence regarding the indiscernibility of the regions in space.  Here, like at other places, Weyl endorsed this Leibnzian argument  from the point of view of ``modern physics'', while adding that for Leibniz this spoke in favour of the unsubstantiality and phenomenality of space and time. On the other hand, for ``real substances'' the Leibnizian monads, indiscernability implied identity.
In this way Weyl indicated, prior to any more technical consideration, that similarity in the Leibnizian sense was the same as {\em objective equality}.  He did not enter  deeper into the metaphysical discussion but insisted that the   issue  discussed in his talk ``is of philosophical significance far beyond its purely geometric aspect''. 

After some remarks on historical shifts of what was considered as ``objective'' properties, e.g. the vertical direction for Democritus or   absolute space for Newton,\footnote{For the last point see also  \cite[100]{Weyl:PMNEnglish}.}
Weyl had good news: ``\ldots 
we can say today in a quite definite manner what the adequate mathematical
instrument is for the formulation of this idea (objectivity, E.S.). It is the notion of group'' (Hs, p. 4).  Weyl did not  claim that this idea solves the epistemological problem of objectivity once and for all,  but at least  it   offers   an {\em adequate mathematical instrument} for the  {\em  formulation} of it. He illustrated the idea in a first step by explaining   the  automorphisms of Euclidean geometry as the structure preserving bijective mappings of the  point set\footnote{Weyl  preferred to avoid the language of sets and used the the terminology of a ``point-field''.}
 underlying a structure satisfying the axioms of 
``Hilbert's classical book on the Foundations of Geometry'' (Hs, p. 4f.). He concluded that for Euclidean geometry these are the {\em similarities},  not the congruences as one might  expect at a first glance (see section \ref{section classical}). In the mathematical sense, we then ``come to {\em interpret objectivity as the invariance under the group of automorphisms}'' (Hs., p.6, emphasis in the original).

But Weyl warned to identify mathematical objectivity with that of natural science, because  once we deal with  {\em real space}  ``neither the axioms nor the basic relations are given'' (Hs, 6). As the latter are extremely difficult to discern, Weyl proposed to turn the tables  and to take the {\em group $\Gamma $ of automorphisms},   rather than the `basic relations' and the corresponding relata, {\em    as the epistemic starting  point}:\footnote{In this context  the ``basic relations'' might be interpreted as   the {\em laws of nature}. For readers who prefer to use the more concrete sounding word {\em symmetries} to ``automorphisms'', Weyl's proposal may be read as the advice to   take the symmetries  as epistemic starting point  and the laws/equations invariant under
this group as  epistemically secondary (derived). Note that Weyl did {\em not} pose the question in terms of which of the two levels is more  {\em fundamental} (see the quote below).}
\begin{quote} 
Hence we come much nearer to the actual state of affairs if we start with the group $\Gamma$ of automorphisms and refrain from making the artificial logical distinction between basic and derived relations. Once the group is known, we know what it means to say of a relation that it is objective, namely invariant with respect to $\Gamma$ \ldots (Hs, 6)
\end{quote}

By  such a well chosen constitutive stipulation it  becomes clear   what objective statements are, although this can be achieved only  at the price  that  ``\ldots we start, as Dante starts in his Divina Comedia, in mezzo del camin'' (Hs. p. 7). A phrase characteristic for  Weyl's later view follows: 
\begin{quote}
It is the common fate of man and his science that we
do not begin at the beginning; we find ourselves somewhere on a road the
 origin and end of which are shrouded in fog (Hs, p. 6).
\end{quote}

Weyl's juxtaposition  of the {\em mathematical} and the {\em physical} concept of {\em objectivity} is worthwhile to reflect upon. The  mathematical objectivity  considered by him is relatively easy to obtain by combining the axiomatic characterization of a mathematical theory (Hilbert) with the epistemic postulate of invariance under a group of automorphisms (Klein). Both are constituted in a series of acts characterized by Weyl in PMN  as {\em symbolic construction}, which is free in several regards. For example, 
the group of automorphisms of Euclidean geometry may be expanded by ``the mathematician'' in rather wide ways (affine, projective, or even ``any group of  transformations'').\footnote{For Weyl  ``any group'' was usually constrained by the condition of differentiability. This was  not necessarily  so for other mathematicians, see e.g. F. Hausdorff's argument for the lack of meaning of an ``absolute'' objective structure of space and time in his philosophical essay {\em Das Chaos in kosmischer Auslese }  in which arbitrary point transformation (bijections) are considered as the most general case \cite{Hausdorff:Chaos}; see \cite{Epple:Hausdorff2006}.}
 In each case  a specific realm of mathematical objectivity is constituted. With the example of the automorphism group $\Gamma$ of  (plane) Euclidean geometry in mind Weyl explained a little later how, through the use of Cartesian coordinates, the automorphisms of Euclidean geometry can be {\em represented} by {\em linear transformations}  ``in terms of reproducible numerical symbols'' (Hs, p.9).

For natural science the situation is quite different; here the freedom of the constitutive act is severely restricted. Weyl described the constraint for the choice of $\Gamma$ at the outset in very general terms:
\begin{quote}
The physicist will question Nature to reveal him her true group of automorphisms (Hs, p. 7).
\end{quote}

This is a striking, even surprising remark.   Different to what a philosopher might expect,  Weyl did not mention, at this place,  the subtle influences induced  by  theoretical evaluations of empirical insights on the constitutive choice of the group of automorphisms for a physical theory. He even did not restrict the consideration to  the range of {\em a} physical theory but aimed at {\em Nature} as a whole. 
Still in 1948/49,  after several  turns of  his own views and radical changes in the fundamental views of theoretical physics, Weyl hoped for an  insight into {\em the  true} group of automorphisms {\em of  Nature} without any further specifications. Of course he did not stop with this general characterization. In the following parts of the talk Weyl explored  in much  more  detail how the ``true group'' of physical automorphisms was shaped with the increasing and deepening empirical and theoretical knowledge of nature.

\section{\small Physical and mathematical automorphisms of classical geometry \label{section classical}}
Looking at classical geometry and mechanics, Weyl followed Newton and Helmholtz in considering  {\em congruence} as the basic relation which lay at the heart of the ``art of measuring'' by the handling of that ``sort of  bodies we call rigid'' (Hs, p. 9). In a short passage he explained how the local congruence relations established by the comparison of rigid bodies can be generalized and abstracted to congruences of the whole space. 
In this respect Weyl followed an empiricist approach to classical physical geometry, based on a theoretical extension of the material practice with {\em rigid bodies} and their {\em motions}. Even the mathematical abstraction to mappings of the whole space carried the mark of their empirical origin and was restricted to the group of  {\em proper  congruences } (orientation preserving isometries of Euclidean space, generated by the translations and rotations) denoted by him as $\Delta^+$. This group  seems to express ``an intrinsic structure of space
 itself; a structure stamped by space upon all the inhabitants of space'' (Hs, p. 10). From a historical perspective,  $\Delta^+$  could serve as the group of physical automorphisms of space until the early 19th century. As we shall see in a moment, Weyl argued that during the  19th century it would be extended to the {\em group of all congruences} $\Delta$ which also includes the orientation reversing isometries (point symmetries, reflections). 

But already on  the  earlier level of physical knowledge, so Weyl argued,  the {\em mathematical automorphisms} of space were  larger than $\Delta$. Even if one sees ``with Newton, in congruence the one and only
basic concept of geometry from which all others derive'' (Hs, p. 10), the group $\Gamma$ of automorphisms in the mathematical sense turns out to be constituted by the {\em similarities}. 

The structural condition for an automorphism $C\in \Gamma$ of classical congruence geometry is that  any pair $(v_1, v_2)$  of congruent geometric configurations is transformed into another pair $(v_1^{\ast}, v_2^{\ast})$ of congruent configurations ($v_j^{\ast}=C(v_j), \; j=1,2$). For evaluating this property Weyl introduced the following diagram:\\

\begindc{\commdiag}[3]
\obj(80,21){$v_1$}
\obj(100,21){$v_2$}
\obj(80,6){$v_1^{\ast}$}
\obj(100,6){$v_2^{\ast}$}
\mor{$v_1$}{$v_2$}{$( T)$}
\mor{$v_1^{\ast}$}{$v_1$}{$\hspace*{5em}(C^{-1})$}
\mor{$v_2$}{$v_2^{\ast}$}{$(C)$}
\enddc

\vspace{1em}
\noindent Because of the condition for automorphisms just mentioned the maps  $C T C^{-1}$ and $C^{-1} T C$ belong to $\Delta^{+}$ whenever $T$ does. By this argument he  showed that the {\em mathematical automorphism group $\Gamma$ is the normalizer of the congruences} $\Delta^+$ in the group of bijective mappings of Euclidean space.\footnote{The same argument is given in \cite[79]{Weyl:PMNEnglish}.}

 This argument contained the {\em mathematical} reason for Weyl's decision in 1918 to consider the similarities as the structure determining morphisms of his  {\em  purely infinitesimal geometry} \cite{Weyl:InfGeo}. More generally, it also explains the reason for his characterization of generalized similarities in his {\em  analysis of the problem of space } in the early 1920s. In 1918 he  translated the relationship between physical equivalences as congruences   to the mathematical automorphisms as the similarities/normalizer of  of the congruences from classical geometry to special relativity (Minkowski space) and ``localized'' them (in the sense of physics), i.e., he transferred the structural relationship to the infinitesimal neighbourhoods of the differentiable manifold characterizing spacetime (in more recent language, to the tangent spaces) and developed what later would be called {\em Weylian manifolds},  a generalization of Riemannian geometry.\footnote{In 1918  Weyl even  hoped that the extension to the similarities was  not only relevant from a mathematical point of view, but  also plays a crucial role for physics (by incorporating electromagnetism into the geometrical structure field and leading to a unified field theory). In his terminology of 1948/49 he had hoped that his generalization of Riemannian geometry to Weylian manifolds of 1918 corresponded to an extension of the physical automorphisms of general relativity. In the transition to the ``new'' quantum mechanics in the 1920s he gave up this hope;  see his corresponding remarks below.} 
In his discussion of the problem of space he generalized the same relationship even further by allowing any (closed) subgroup of the general linear group as a candidate for characterizing generalized congruences at every point. 

Moreover, Weyl  argued that the enlargement of the physico-geometrical automorphisms of classical geometry (proper congruences) by the mathematical automorphisms (similarities)  sheds light on {\em Kant's riddle} of the  ``incongruous counterparts''. Weyl presented it   as the question: Why are ``incongruous counterparts'' like the left and right hands  intrinsically indiscernible, although they cannot be transformed into another by a proper motion?  From his point of view the intrinsic indiscernibility could be  characterized by the mathematical automorphisms $\Gamma$. Of course,  the congruences $\Delta$ including the reflections are part of the latter,  $\Delta \subset \Gamma$; this implies indiscernibility between ``left and right'' as a special case. In this way  Kant's  riddle was solved by a Leibnizian type of argument. Weyl very cautiously indicated a philosophical implication of this observation
\begin{quote}
And he (Kant, E.S.) is inclined to think that only 
transcendental idealism is able to solve this riddle. No doubt, the meaning of
congruence and similarity is founded in spatial intuition. Kant seems to aim
at some subtler point. But just this point is one which can be completely
clarified by general concepts, namely by subsuming it under the general and
typical group-theoretic situation explained before \ldots (Hs. p. 7).
\end{quote}
Weyl stopped here without discussing the relationship between group theoretical methods and the ``subtler point'' Kant aimed at more explicitly. But we may read this remark as an   indication that he considered his reflections on  automorphism groups as a contribution to the  transcendental analysis of the conceptual constitution of modern science. For Weyl this meant, of course, modern science in the sense of the 20th century, i.e., taking general relativity and quantum physics into account. A little later, in his book on {\em Symmetry},  he  went a tiny step further. Still with the Weylian restraint regarding the discussion of  philosophical  principles he stated: ``As far as I see all a priori statements in physics have their origin in symmetry'' \cite[126]{Weyl:Symmetry}.

To prepare for the following, Weyl  specified the subgroup $\Delta_o \subset \Delta$ with all those transformations that fix one point ($\Delta_o = O(3,\R)$, the orthogonal group in 3 dimensions, $\R$ the field of real numbers). In passing he remarked:
\begin{quote}
 In the four-dimensional world the Lorentz group takes the place of the orthogonal
group. But here I shall restrict myself to the three-dimensional 
 space, only occasionally pointing to the modifications, the inclusion of time
into the four-dimensional world brings about. (Hs, p. 13f.)
\end{quote}

Keeping this caveat in mind (restriction to three-dimensional 
 space) Weyl characterized the ''{\em group of automorphisms of the physical world}'', in the sense of classical physics (including  quantum mechanics) by the combination (more technically, the semidirect product $\rtimes$) of translations and rotations, while the mathematical automorphisms arise from a normal extension:
\begin{itemize}
\item[--] physical automorphisms $\Delta \cong \R^3 \rtimes \Delta_o$ with $\Delta_o \cong O(3)  $, \\ respectively $\Delta \cong \R^4 \rtimes \Delta_o$  for the Lorentz group $\Delta_o \cong O(1,3)$,
\item[--] mathematical automorphisms $\Gamma =\R^+ \times \Delta$ \\($\R^+$ the positive real numbers with multiplication).
\end{itemize}

In Weyl's view  the {\em difference between mathematical and physical automorphisms } established a fundamental distinction between mathematical geometry and physics.
\begin{quote}
{\em Congruence, or physical equivalence}, is a geometric concept, the meaning of which refers to the 
laws of physical phenomena; the congruence group $\Delta$ is essentially 
the group of physical automorphisms. If we interpret geometry as 
an abstract science dealing with such relations and such relations 
only as can be logically defined in terms of the one concept of congruence, then the group of {\em geometric automorphisms} is the normalizer 
of $\Delta$ and hence {\em wider} than $\Delta$. (Hs, p. 16f., emphasis E.S.)
\end{quote}
 He considered this as a striking argument against  what he considered to be the Cartesian program of a reductionist  geometrization of physics (physics as the science of {\em res extensa}):
\begin{quote}
According to this conception, Descartes's program of reducing physics to geometry would involve a
vicious circle, and the fact that the group of geometric automorphisms is wider than that of physical automorphisms would show that 
such a reduction is actually impossible.'' (Hs, p. 16f., similar in \cite[83]{Weyl:PMNEnglish})
\end{quote}
In this Weyl alluded to an illusion he himself had shared for a short time as a young scientist.
After the creation of his gauge geometry in 1918 and the proposal of a geometrically unified field theory of electromagnetism and gravity he believed, for a short while,  to have  achieved  a complete geometrization of physics.\footnote{In the third German edition of {\em Space - Time - Matter} (1919) Weyl included his  proposal for a unified field and matter theory. At the end of the book he drew the following conclusion:
\begin{quote} We have realized that physics and geometry collaps to one, that the world metric is  a physical reality, and even the only one. In the end, the whole physical reality appears to be a mere form; not geometry has become physicalized, but  physics has turned into geometry. \cite[263, my translation, E.S.]{Weyl:RZM3}
\end{quote}
(``Wir hatten erkannt, da\ss{}  Physik und
Geometrie schlie\ss{}lich zusammenfallen, da\ss{}  die Weltmetrik eine, ja vielmehr
die physikalische Realit\"at ist. Aber letzten Endes erscheint so diese
ganze physikalische Realit\"at doch als eine blo\ss{}e Form; nicht die Geometrie
ist zur Physik, sondern die Physik zur Geometrie geworden.'')\\
In the next years, and already in the following editions of {\em Space - Time - Matter},  Weyl  withdrew  step by step from this geometrization of physics perspective. In his Rouse Ball lecture 1930 he likened it with   premature  ``geometrical jumps into the air (geometrische Luftspr\"unge)'' which had lost contact with  the ``solid ground of physical facts'' (of quantum physical observations) \cite[343]{Weyl:Rouse_Ball}.}

He gave up this illusion in the middle of the 1920s under the impression of the  rising  quantum mechanics. In his own contribution to the new quantum mechanics groups and their linear representations played a crucial role. In this respect the mathematical automorphisms of geometry and the physical automorphisms ``of Nature'', or more precisely the automorphisms of physical systems,  moved even further apart, because now the physical automorphism started to take non-geometrical material degrees of freedom into account (phase symmetry of wave functions and, already earlier, the permutation symmetries of $n$-particle systems). 

But already during the 19th century the physical automorphism group had acquired a  far deeper aspect  than that of the mobility of rigid bodies: 
\begin{quote} In physics we have to consider not only points
but many types of physical quantities such as velocity, force, electromagnetic field strength, etc. \ldots
\end{quote}
 All these quantities can be represented, relative to a Cartesian frame,
 by sets of numbers such that {\em any orthogonal transformation} $T$ performed on the coordinates keeps the basic physical relations, the physical laws, invariant. Weyl accordingly stated:
\begin{quote}
{\em All the laws of nature are invariant under the transformations thus
induced by the group $\Delta$.} Thus physical relativity can be completely described
by means of a group of transformations of space-points. (Hs. p. 14, emphasis in orginal)
\end{quote}

By this argumentation Weyl described a deep shift which ocurred in the late 19th century for  the understanding of physics. He  described it as an {\em extension of the group of physical automorphisms}. The laws of physics (``basic relations'' in his more abstract terminology above) could no longer  be directly  characterized by the motion of rigid bodies because the  physics of fields, in particular of electric and magnetic fields, had become central. In this context, the motions of material bodies lost their epistemological primary status and the physical automorphisms  acquired a more abstract character, although they were still   completely characterizable in geometric terms, by the full group of Euclidean isometries. The indistinguishability of left and right, observed already in  clear terms by Kant, acquired the status of a physical symmetry in electromagnetism and in crystallography.\footnote{In geometrical crystallography the point inversion symmetry played a crucial role already a short time after Kant's death  and was developed during the 19th century in both, the atomistic and the dynamistic, programs of crystallography. With the transition to group theoretic descriptions of crystallographic symmetries the extension of proper motions by orientation reversing isometries was made explicit between Camille Jordan's paper of 1869 {\em M\'emoire sur les groupes de mouvements} and Sohncke's adaptation  in crystallographic studies to Schoenflies' papers on crystallographic groups in  1888ff. under the influence of F. Klein \cite[chap. I]{Scholz:Habil}. }

Weyl thus insisted that in classical physics the physical automorphisms could be  characterized by the group $\Delta$ of Euclidean isometries, larger than the physical congruences (proper motions) $\Delta^+$ but smaller than the mathematical automorphisms (similarities) $\Gamma$.\footnote{Similar arguments can be found  in \cite[83]{Weyl:PMNEnglish} and \cite[129]{Weyl:Symmetry}. The argument was apparently meant to hold, mutatis mutandis, also for special relativistic physics (``\ldots only occasionally pointing to the modifications, the inclusion of time
into the four-dimensional world brings about'').  }

This view fitted well to  insights which Weyl drew from recent developments in quantum physics. He  insisted  --  differently to what he had thought  in 1918 -- on  the  consequence that ``length is not relative but absolute'' (Hs, p. 15). He argued that {\em physical length} measurements were  no longer dependent on an arbitrary chosen unit, like in Euclidean geometry. An ``absolute standard of length''  could  be fixed by the quantum mechanical laws of the  atomic shell:
\begin{quote} The atomic constants of charge and mass of the electron 
atomic constants and 
Planck's quantum of action $h$, which enter the universal field laws 
of nature, fix an {\em absolute standard of length}, that through the {\em wave
lengths of spectral lines} is made available for practical measurements. (Hs, 15, emphasis E.S.)
\end{quote}

This statement was important for Weyl; he repeated the passage  in  \cite[83]{Weyl:PMNEnglish} and \cite[129]{Weyl:Symmetry} in similar words. It demarcates a crucial difference of Weyl's mature view of the physical metric from his earlier ones (1918 until about 1924). In the terminology of his 1918 debate with Einstein, Weyl came now to accept that the laws of quantum physics and the constitution of the atom establish a kind of ``universal bureau of standards (Eichamt)'', contrary to what pure field physics made him expect in 1918.\footnote{In this context it is  interesting to see that the International System of Standards (SI) is  now substituting  the pragmatic convention of the Paris {\em urmeter} by implementing  a fixed standard of length on the basis of the ground state hyperfine splitting frequency of the caesium 133 atom and reference to a collection of natural constants (velocity of light $c$, Planck constant $\hbar$, elementary charge $e$, Boltzmann, Avogadro and Rydberg constants).  }

\section{\small The 'shock of relativity' \label{section shock}}
``So far so good. But now comes the shock of general relativity theory. It taught us that the group of physical automorphisms is much larger than we had assumed so far \ldots'' 
 (Hs, p. 16).  With these words Weyl turned towards the shift general relativity  brought about for the understanding of mathematical and physical  automorphisms of geometry, or even ``Nature'' as such. 
He described the mathematical structure of differentiable and Riemannian manifolds by means of coordinate systems and characterized the tangent spaces by point dependent ``Cartesian'' (orthonormal) vector bases.
This description of orthonormal frames was a simplification of a method for generalizing Dirac's electron theory to the general relativistic context \cite{Weyl:1929Dirac}.\footnote{Similarly in \cite[88]{Weyl:PMNEnglish}, while the following discussion of physical automorphisms is much more detailed in the talk (Hs, pp. 16ff.).}  

Weyl described the automorphism group of general relativity verbally, i.e., without formulae, as containing  ``all transformations (satisfying
certain continuity or differentiability restrictions)'' (Hs, p. 16), i.e., the diffeomorphisms of the space(-time) manifold $M$. A little later he characterized this part of the physical automorphism group (sloppily) by its expression in coordinates  and talked about
\begin{quote}
\ldots \ldots the group $\Omega$  of all coordinate transformations,
which expresses the generally relativistic molluscous nature of space as the
`field of possible coincidences' (Hs, p. 18).
\end{quote}
In addition, the different choices of orthonormal frames at each point of the manifold are associated  to point dependent ``rotations'' of type $\Delta_o$ (orthogonal, or Lorentz transformations). They  have to be taken into account for transforming between different pictures of a general relativistic field constellation. Thus:
\begin{quote}
The laws of nature are independent
of the arbitrariness involved in these two acts. In other words, they are 
invariant (1) with respect to arbitrary continuous (or rather differentiable)
coordinate transformations, (2) with respect to any rotation of the Cartesian
frame at $P$, a rotation that may depend in an arbitrary manner on the point
$P$. (Hs, p. 18)
\end{quote}

On a first reading two, or even three, features of Weyl's characterization of the repercussions of the ``shock of relativity'' on the perception of the physical automorphisms may appear puzzling.  He did not specify, at least not  in general terms,  what the ``independence of the arbitrariness'' under the respective choices of coordinates and frames precisely  meant for the natural laws. At a superficial glance his characterization may seem  to be subject to Kretschmann's criticism of the meaning of general covariance for general relativity. But  Weyl made it quite clear that he understood  the {\em invariance of the laws of nature} under the physical automorphisms in a {\em  strong sense}, i.e, without assuming an additional background structure (e.g. the Euclidean metric in the case of Newtonian mechanics) which would have to be transformed concomitantly  with the dynamical quantities appearing in the natural laws.\footnote{This difference   is sometimes  considered as the crucial difference between {\em symmetries} of the laws and {\em covariance of the equations} of motion, see \cite{Giulini:symmetry} or the  commentary by the same author in the 3rd edition of \cite{Weyl:Symmetrie}. }  
Following Einstein in this regard, Weyl expressed this idea for general relativity in  clear terms:
\begin{quote}
The metric structure, and the inertial structure derived from it, exert a powerful influence upon
all physical phenomena. But what acts must also suffer. In other words, the
metric structure must be conceived as something variable, like matter and
like the electromagnetic field, which stands with all other physical quantities
 in the commerce of interaction: it acts and suffers reactions. Only by admitting
the metrical field as a variable physical entity among the other physical
quantities can the principle of general relativity be carried through. (Hs, p. 16)
\end{quote}

The second puzzling feature is Weyl's rather generous quid pro quo of differentiable coordinate transformations and diffeomorphisms of the manifold, mentioned already above. But this was a general way of expressing diffeomorphisms by Weyl. He did not like transfinite sets, and therefore tended to avoid the description of manifolds by  locally Euclidean Hausdorff spaces endowed with what later would be called an atlas of coordinate systems. He rather preferred a definition by equivalences of coordinate domains, because he considered this  a more constructive approach to the concept of manifold. This being said, we need not bother much about this peculiar mode of expression in our context.

The third puzzling feature relates to Weyl's treatment of the localized, point dependent operation of the (Lorentz) orthogonal group on the tangent spaces. This aspect deserves more attention. 

\section{\small Automorphisms of general relativity as a gauge group \label{section gauge group}}
 Weyl dealt her with what  later would be called a {\em gauge group} or, more precisely,  a {\em gauge automorphism group} over spacetime $M$ with structure group $G=\Delta_o$. The situation is complicated by different uses of the term {\em gauge group} in the present literature. In fibre bundle language, referring to a principal fibre bundle $G \lefttorightarrow P\rightarrow M$  over the base manifold $M$ with structure group $G$, some authors consider the group  $\mathcal{G}_P$ of fibrewise (``vertically'') operating bundle automorphisms as the gauge group of $P$. The elements of  $\mathcal{G}_P$ are called (global) {\em gauge transformations}.

In another, more general, view  the group $\mathcal{G}(P)$ of all (equivariant) bundle automorphisms is considered as the gauge group of $P$. In order to disambiguate the terminology, the elements of $\mathcal{G}(P)$ might better be called (bundle) {\em automorphisms},\footnote{This is the terminology in, e.g., \cite[46]{Bleecker:Gauge_theory}. I thank an anonymous referee for the literature hint.}
and  $\mathcal{G}(P)$  the {\em  group of gauge automorphisms}. 
 Different from the first definition, the elements of $\mathcal{G}(P)$ allow diffeomorphic transformations of the base. In our context the latter definition is more appropriate, because it expresses Weyl's description of the automorphism group of general relativity in modernized terms. 

To make things even more involved, one often speaks of  gauge transformations for describing the local {\em changes of trivialization} of a given fibre bundle, including their operation on the local representatives of connections. This last reading of gauge transformations corresponds  to  a local  realizations of the vertical gauge transformations in the sense of  $\mathcal{G}_P$, just  like  the  differentiable coordinate transformations correspond  to diffeomorphisms of a manifold.

The physical automorphism group of general relativity, intuitively described by  Weyl as composed  by two components ($\Omega$ and the family of point dependent $\Delta_o$-s) would have been difficult to formulate at the end of the 1940s.
But already a few years later,   in the early 1950s, the next generation of mathematicians learned to describe such a group $\mathcal{G}(P)$ by a principle fibre bundle $P$ over $M$, $P \longrightarrow M$ with structure group $G\cong \Delta_o$.\footnote{Norman Steenrod, since 1947 at the faculty of Princeton University, and the Strasbourg group of differential geometers about Charles Ehresmann played a crucial role for this development. Ehresmann was a student of E. Cartan and had been in close contact with Weyl during his time at G\"ottingen (1930/1931) and again   in Princeton from 1932 to 1934 \cite{Zisman:1999}.} 
The latter describes both, the fibres of $P$ and their allowable transformations ($G$ operates fibrewise on $P$). As  $\mathcal{G}(P)$ consists of the diffeomorphisms of the bundle $P$, which induce diffeomorphisms on the base manifold $M$ and map the fibres in such a way that the group operations are respected, it formalizes Weyl's intuition of the liberty of choice of a local reference system (orthogonal frame) at every point quite well.\footnote{After a choice of  a local frame a numerical representation of the (Lorentz) orthogonal  is specified, and the operation of the orthogonal group is ``trivialized''. A change of frames leads to another representation arising from the first one by conjugation (in modernized terminology a local change of trivialization).  }
An element of  $\mathcal{G}(P)$ induces  a local diffeomorphism of the base manifold $M$   and, moreover, it specifies an orthogonal transformation at each point of $M$. This   corresponds to a point dependent change  of frames in Weyl's description. Therefore  $\mathcal{G}(P)$ is a well adapted modernized (and only minimally anachronistic)  global expression for Weyl's automorphism group of general relativity. 

Weyl compared the new   group  with the automorphisms of special relativity or even classical physics. The translations of the classical automorphism group were generalized by the diffeomorphism group $\Omega$ and the ``rotations'' (Lorentz transformations) became point dependent. 

But this was only a first step into modernity,  not yet the final word. For the generalization of the Dirac theory to general relativity  it turned out to be necessary (and in Weyl's view also natural) to extend the structure group $\Delta_o$ by a complex phase factor to $\tilde{\Delta}_o \cong \Delta_o\times U(1)$. Weyl indicated this step by the transition to a (2-component) spinor representation of the rotation group (resp. Lorentz group) and  added that the latter was underdetermined by a complex factor of norm 1:
\begin{quote}
The two components $\psi_1, \psi_2$  of the electronic wave field have the peculiarity
 that they are determined by the local frame $\mathbf{f}$ only up to an arbitrary factor $\alpha = e^{i \lambda}$ of modulus 1. The real $\lambda$ could be described as a common shift of
phase in the two complex quantities $\psi_1, \psi_2$. This gauge factor $\alpha$ adds one
more parameter to the representing group of transformations \ldots (Hs, p. 20)
\end{quote}
This extension leads to a general relativistic theory of the electron field, which had been proposed independently by Weyl and V. Fock in 1929.\footnote{\cite{Vizgin:UFT,Straumann:DMV,Scholz:Fock/Weyl}.} In modernized description Weyl finally considered  the  gauge automorphisms $\mathcal{G}(P)$  
of  a bundle 
\beq   \Delta_o\times U(1)  \cong \tilde{\Delta}_o \lefttorightarrow P\rightarrow M \label{G(P)}
\eeq 
constructed over spacetime $M$ (considered as a Lorentz manifold) by extending the  orthogonal frame bundle as the {\em physical automorphisms} of (phase extended) general relativity. 

He indicated that the new automorphism group corresponds to important {\em conservation principles} and/or {\em structure theorems} of physics: conservation of energy and momentum correspond to the coordinate transformations of spacetime, symmetry of the energy tensor to the point-dependent ``rotations'' (HS, p. 19). The first part of this remark should not be taken literally; it is (overly) simplified. While the second part of the statement (symmetry of energy tensor)  is a consequence of the rotational symmetries of the automorphism group, the conservation of energy/momentum is more problematic. It makes sense in special relativity and under restriction to the Poincar\'e group, or more generally under strong homogeneity conditions of spacetime and a restriction to adapted frames of reference; in the general case the conserved quantities exist mathematically (Noether currents) but do not allow a coordinate and observer independent physical interpretation.   Weyl had discussed this point more precisely in his 1929-paper; here he simplified it, probably for didactical reasons, a bit too strongly.\footnote{Weyl  touched the problematics of the Noether theorems, without quoting Noether.  For establishment, reception and philosophical discussion of the Noether theorems see  \cite{Kosmann:Noether,Rowe:Noether_thms,Brading:energy_conservation,Sus:conservation}. \label{fn Noether}}

The justification of the attribute ``physical'' for the diffeomorphism component of the automorphism group of general relativity is therefore more subtle than  admitted here by Weyl. But he had  given another necessary criterion for their {\em physicality} earlier  (Hs. p. 16): the invariance of the laws of nature, which included the invariance of the Lagrangian density for a  field theory like Einstein gravity, {\em without any covariant non-dynamical background structure}  (see above) -- a subtle characterization of physicality indeed. 

The simplified discussion of conservation of energy/momentum  made it easier to emphasize the physcial role of  the phase gauge invariance of electromagnetism. After the rather formal introduction of the phase extension of the structure group   $\tilde{\Delta}_o \cong \Delta_o\times U(1)$, Weyl announced:
\begin{quote}
The law of conservation of charge corresponds to it [phase invariance, E.S.] 
in the same manner as the law of the conservation of 
energy-momentum corresponds to the invariance under coordinate 
transformations. (Hs, p. 20)
\end{quote}
This feature was of particular importance for Weyl's 1929 extension of the automorphism group of general relativity. It is, in fact, less problematic than the energy conservation statement mentioned above.\footnote{The reason lies in the specific structure of the phase gauge extension; see \cite{Brading:Noether_Weyl}.}

On the other hand, the ``gauging'' of phase (the choice of a local trivialization in mathematical language) relied on a completely abstract, symbolical choice; the same holds for the  localized phase transformations. While gauging  the  (Lorentz) orthogonal group could still be understood as a choice of frames, i.e.  as a result of a choice of  observer systems with point dependent relative motions, and the gauging of Weyl's 1918 scale group could be understood as as point dependent choice of units of measurement, the gauging of phase was emptied of any direct empirical content, it became ``descriptive fluff''  \cite{Earman:symmetrybreaking}. This  poses the question in which respect Weyl's talk of ``physical automorphisms''  deals with more than a bunch of transformations of descriptive fluff.

The answer was already indicated  by Weyl by emphasizing the {\em invariance of physical laws} (without non-dynamical covariant background structure) and the reference to {\em conserved quantities} or  {\em structural consequences} of what would later  become known as conserved Noether currents. In the language of fibre bundles we may rephrase Weyl's argument by calling  to attention that the gauge automorphism group $\mathcal{G}(P)$ operates on the whole system of dynamical variables in a way which allows to deal with the invariance of the ``laws of nature'' (in particular the Lagrangian densities in case of Lagrangian field theories) in a mathematically precise way without introducing new non-dynamical background features. This  allows to decide whether a given group consists of automorphisms  of the theory and gives a necessary condition for the latter's ``physicality'' (background independence); but it  does not yet give  a sufficient criterion. We may still be able to extend the bundle of a given  physically meaningful structure by a new symmetry {\em and} a hypothetical new dynamical field which  avoids the appearance of a non-dynamical background structure.\footnote{A good example is the reintroduction of Weyl's scale symmetry into gravity by Utiyma and Dirac in the early 1970s. They added  a hypothetical scale covariant scalar field coupled to the Hilbert term, similar to the Jordan-Brans-Dicke theory, and endowed it with a kinetic term of its own (Weyl geometric scalar tensor theory).} 
The crucial question is as to whether or not the extension of the {\em structure as a whole},  not every single field in isolation, leads to {\em new physical insight}. This may be of any kind, conservation laws are only the most prominent example.

Weyl's proposal of 1918 for introducing a metrical (point-dependent) scale gauge invariance as a fundamental symmetry of electromagnetism was substituted by a point-dependent phase choice. He discussed the transition in some detail and  emphasized that in the ``old theory'' the unit of the electromagnetic potential was related to Einstein's cosmological constant, while in the new theory the potential was ``measured not in an unknown cosmological, but in a known atomic unit'' (Hs, p. 21). Moreover, quantum theory seemed to speak for the existence of an ``absolute standard of length'' (by fixing the frequency of atomic clocks, see above); he   therefore concluded:

\begin{quote}
I have no doubt that  my old speculative theory
has to be given up in favor of the quantum 
mechanical principle of phase invariance, that rests on sound empirical foundations. The facts of atomism teach us that length is 
not relative but absolute, and that the origin for the standard of 
length must be sought not in the cosmos as a whole, but in the 
elementary material particles. The additional group parameter is 
not geometric dilatation, but electronic phase shift. I was on the 
right track in 1918 as far as the formalism of gauge invariance is 
concerned. But the $\psi$'s on which to hang the gauge factor $\alpha$ were 
utterly unknown at that time, and so I wrongly hitched it on to 
Einstein's gravitational potentials $g_{ik}$. (Hs, p. 21f.)
\end{quote}
Weyl definitely no longer considered the scale invariance of 1918 to be  of physical relevance. After 1929 he saw  its status  reduced to being  part of the mathematical automorphisms of general relativity, while the physical automorphisms were extended to  include a ``gauged'' (point-dependent) phase symmetry. 

\section{\small Weyl's distinction  in the context of gauge structures \label{section Weyls distinction}}
In this passage Weyl  attributed different epistemic qualities to two possible extensions $\tilde{\Delta}_o = \Delta_o\times  H$ of the structure group (here $\Delta_o$  the  orthogonal group)  by  abelian factors  $H=U(1)$ (phase) or $H=\R^+$ (scale). The distinction sheds light on the central topic  of the talk because  the first one lead to an extension of the {\em physical automorphisms} of general relativity, while the second one appeared as an extension of the {\em mathematical} automorphisms only. This was a nice analogy to the relationship between {\em congruences} and {\em similarities} in the classical case. But Weyl avoided to give an explicit and general description of the  demarcation between the  physical and the mathematical automorphisms of ``Nature''.

 On the other hand,  Weyl  developed different aspects and criteria for the distinction he was after in his  case by case discussion of  three different phases in the development of modern physics (early modern, late 19th century, early 20th century). In the following we  try   to distill the principles underlying the case-bound criteria which served Weyl for his distinction   in   in a  more general and explicit  form. Then we can see whether they tell us  something about other theoretical developments, partly contemporary to Weyl but not dicussed  in his talk, like Einstein-Cartan gravity, but also about later  ones  like Sciama-Kibble  and Utiyma-Dirac gravity. Maybe they even can enrich the  philosophical debate on the  gauge theories of the standard model of elementary particle physics.

I propose the following generalization and  interpretation:   In principle
  any  normal extension $\mathfrak{A}_{m}$ of a given automorphism group  $\mathfrak{A}_{o}$ may constitute a new level of {\em mathematical objectivity}  for ``the mathematician'', if it leaves some interesting structure invariant.  For a theory aiming at {\em physical objectivity}, on the other hand,  the automorphisms have to be constrained to the largest subgroup  $\mathfrak{A}_{p} \subset \mathfrak{A}_{m}$ which   satisfies  the following criteria:
\begin{itemize}
	\item[(i)]  Basic physical relations (``laws of nature'') are invariant under $\mathfrak{A}_p$.
	\item[(ii)] There are no non-dynamical background structures (no ``absolute'' elements).
	\item[(iii)] All degrees of freedom of $\mathfrak{A}_p$ have some physically meaningful, perhaps even striking, consequence for the theory as a whole. Such consequences may consist in a crucial heuristic role of $\mathfrak{A}_p$ for determining   the principles underlying the ``laws of nature'', e.g. in decisive constraints for the Lagrange density.
\end{itemize}
Criterion (iii) has been  vaguely formulated,  because it may be realized by quite different types of consequences. The most prominent  ones, beside the symmetry constraint for the Lagrangian, are the Noether equations  of the respective symmetries which may lead to empirically relevant conserved quantities  ({\em Noether charge paradigm}). But this type is not the only one;  there  may be other consequences of a more {\em structural  nature}.

In Weyl's discussion this was  the case  for the rotational degrees of freedom of his physical automorphism (\ref{G(P)}); these implied  the symmetry of the energy-momentum tensor rather than a conserved charge.
In addition, Weyl's argument for the physical character of the diffeomorphism component in the base manifold of (\ref{G(P)})  by hinting at the conservation of energy and momentum has to be taken {\em cum grano salis}, because the conserved Noether charges of the diffeomorphism degrees of freedom cannot be given an empirically relevant meaning without assuming special conditions, e.g. asymptotic flatness (cf. footnote \ref{fn Noether}). On the other hand, the postulate of diffeomorphism invariance constrains  the choice of the Lagrangian for ``the metric structure, and the inertial structure derived from it''   (Hs, p. 16, quote above); it leads to the Hilbert action plus a constant term which may, but need not, vanish. Therefore the diffeomorphisms are clearly part of the physical automorphism group of general relativity.

\section{\small A side glance at other gauge theories \label{section side glance}}

From a more recent point of view, we may add that in the standard model of elementary particle physics the most important  physical insight of the underlying gauge structures, in addition to the symmetry constraint for the Lagrangians,  consists  in  the  property of {\em  renormalizability}.  The  Noether equations do not lead to empirically relevant conserved quantities;  but  under quantization they develop a structural effect   (``Slavnov-Taylor identies'')  which is  crucial  for the renormalizability of the theories. They thus establish  a crucial precondition for the empirical relevance of the theory.   Weyl often insisted on the necessity for comparing a theory with the respective segment of ``the  world'' only as a whole. For him, the ``physicality'' of automorphisms could   be established by features of a much more general nature than one might expect from a strictly  empiricist point of view. 
Perhaps his conception may shed light on the  discussion in the present literature on philosophy of physics   why, or even whether,  gauge transformations may be of physical significance, if taking into account that gauge transformations seem to deal with nothing more than  ``descriptive fluff''. 

Before we come to the specific Weylian point, we have to remember that  gauge transformations appear of primarily  descriptive nature only if we consider them in their function as changes of local (in the mathematical sense) changes of trivializations. In this function they are comparable to the transformations of the coordinates in  a differentiable manifold, which  also seem to have a purely ``descriptive'' function. But  the coordinate changes stand in close relation to (local) diffeomorphisms, like in  Weyl's argumentation above. Therefore  the postulate  of coordinate independence of natural laws, or of the Lagrangian density,  can and is being restated in terms of diffeomorphism invariance in general relativity. Similarly, the  local changes of trivializations may be read as local descriptions of elements of $\mathfrak{G}_P$ in the notation above, i.e., as fibrewise operations of  gauge automorphisms.   $\mathfrak{G}_P$ is a subgroup of the more general gauge automorphism group  $\mathfrak{G}(P)$ which includes transformations of the base like in Weyl's discussion; it  thus reflects an important part of the structural features of the bundle $G\lefttorightarrow P \rightarrow M$. 

The question as to whether or not the  automorphisms of $\mathfrak{G}_P$, or even of  $\mathfrak{G}(P)$,  express crucial  physical properties (item (iii) above) has nothing to do with the specific gauge nature of the groups, but hinges on the more overarching question of physical adequateness and physical content of the theory. The question of whether or why gauge symmetries can express physical content is  not much different from the Kretschmann question of whether or why coordinate invariance of the laws, respectively coordinate covariance description of a physical theory, can have physical content. In the latter case the answer to the question has been dealt with  in the philosophy of physics literature in great  detail. Weyl's contribution to both levels of the debate, the original Kretschmann question {\em and} the gauge symmetry question has been described above; his answer is contained in his thoughts on the distinction of physical and mathematical automorphisms. In our rephrasing they are basically covered by (iii).

Let us shed a side-glance at gravitational gauge theories not taken into account by Weyl in his talk.  In Einstein-Cartan gravity, which later turned out to be equivalent to Kibble-Sciama gravity,   the localized rotational degrees of freedom  lead to a conserved spin current and a non-symmetric energy tensor.\footnote{\cite{Trautman:EC,Hehl:Dennis}.} This is a structurally pleasing  effect, fitting roughly into the Noether charge paradigm, although with a  peculiar ``crossover'' of the two Noether currents and the currents feeding the r.h.s of the dynamical equations, inherited from Einstein gravity and Cartan's identification of {\em  translational} curvature with {\em torsion}. The rotational  current, spin, feeds the dynamical equation of translational curvature;  the translational current, energy-momentum, feeds the rotational curvature in the (generalized) Einstein equation.\footnote{\cite[sec. 9, 10]{Hehl:Dennis}}
 According to the experts it may acquire physical relevance only if energy densities surpass the order of magnitude $10^{38}$  times the density of neutron stars.\footnote{\cite[p. 108]{Blagojevic/Hehl}, \cite[194]{Trautman:EC}.}
By this reason  the current cannot yet be considered a physically striking effect. 
It may turn into one, if  gravitational fields corresponding to extremely high energy densities acquire empirical relevance. For the time being, the rotational current can  safely be neglected, Einstein-Cartan gravity reduces effectively to Einstein gravity,  and Weyl's argument for the symmetry of the energy-momentum tensor remains the most  ``striking consequence'' in the sense of (iii) for the rotational degrees of freedom. 

On the other hand, the translational degrees of freedom give a more direct expression for the Noether currents of energy-momentum than the diffeomorphisms. The physical consequences  for the diffeomorphism degrees of freedom reduce to the invariance constraint for the Lagrangian density for Einstein gravity considered as a special case of the Einstein-Cartan theory (with effectively vanishing spin). 
Besides these minor shifts,  it may be 
 more interesting   to realize  that   the approach of Kibble and Sciama agreed nicely  with Weyl's methodological remark that for understanding nature we  better ``start with the group $ \Gamma$ of automorphisms and refrain from making the artificial logical distinction between basic and derived relations \ldots'' (Hs. 6, see above, sec. 1). This describes quite well  what Sciama and Kibble did. They started to explore the consequences of  localizing  (in the physical sense)  the translational and rotational degrees of freedom of special relativity. Their theory was built around the generalized automorphism group arising from localizing the Poincar\'e group.

\section{\small The enduring role of the Lorentz group \label{section Lorentz group}}
At the end of his talk Weyl pondered on the reasons why the structure group of the physical automorphisms  still contained the ``Euclidean rotation group'' (respectively the Lorentz group, E.S.)  in such a prominent role:
\begin{quote}
The Euclidean group of rotations has survived even such radical 
changes of our concepts of the physical world as general relativity 
and quantum theory. What then are the peculiar merits of this group 
to which it owes its elevation to the basic group pattern of the 
universe?  For what `sufficient reasons' did the Creator choose  this group and no other?'' (Hs, 22)
\end{quote}

He reminded his audience that Helmholtz had  characterized $\Delta_o \cong  SO(3,\R)$) by the ``fact that it gives to a rotating solid what we may call its just degrees of freedom'' of a rotating solid body;\footnote{Weyl explained  this metaphorical description by  what is now being called {\em simple flag transitivity}: ``any incident set of $1-, \, 2-, \, \ldots, (n-1)-$ dimensional directions can be carried into any other such set by a suitable
but uniquely determined element of the group'' (Hs. p. 22).}
 but this method ``breaks down for the Lorentz 
group that in the four-dimensional world takes the place of the orthogonal group in 3-space'' (Hs, p. 22). In the early 1920s he himself had given another characterization living up to the new demands of the theories of relativity in his mathematical analysis of the problem of space.\footnote{ ``I have given another characterization free 
from this blemish by showing that the group of linear transformations 
that leave a non-degenerate quadratic form invariant is the only one 
that ties affine connection to metric in the manner so characteristic 
for Riemannian geometry and Einsteinian gravitation.'' (Hs, p. 22)
Cf. \cite{Bernard:Barcelona,Scholz:2015WeylCartan}}

But now, twenty years later, he wanted to go  further. A bit earlier in his talk he mentioned the idea that the Lorentz group might play its prominent role for the physical automorphisms because it expresses deep lying matter structures; but he  strongly qualified  the idea immediately after having stated it:
\begin{quote}
Since 
we have the dualism of invariance with respect to two groups and $\Omega$ 
certainly refers to the manifold of space points, it is a tempting 
idea to ascribe $\Delta_o$ to {\em matter} and see in it a characteristic of the
localizable elementary particles of matter. I leave it undecided 
whether this idea, the very meaning of which is somewhat vague, has 
any real merits. (Hs, p. 19)
\end{quote}

Coming closer to the end of his talk he indicated another, rather more mathematical idea, even more abstract than his approach in the mathematical analysis of the space problem
\begin{quote}
\ldots  But instead of 
analysing the structure of the orthogonal group of transformations 
$\Delta_o$,   it may be wiser to look for a characterization of the group $\Delta_o$ 
as an {\em abstract} group. Here we know that the homogeneous n-dimensional orthogonal groups form one of 3 great classes of {\em simple Lie 
groups}. This is at least a partial solution of the problem. (Hs, 22)
\end{quote}
He left it open why it ought to be ``wiser'' to look for abstract structure properties in order to answer a natural philosophical question. Could it be that he wanted to indicate an open-mindedness toward the  more  structuralist perspective  on  automorphism groups, preferred by the young algebraists around him at Princetion in the 1930/40s? Today the classification of simple Lie groups distinguishes 4 series, $A_k, B_k, C_k, D_k$. 
  Weyl  apparently counted the two orthogonal series $B_k$ and $D_k$ as one. The special orthogonal groups in even complex space dimension form the series   of simple Lie groups of type $D_k$, with complex form $SO(2k,\C)$ and real compact form $SO(2k,\R)$. The special orthogonal group in odd space dimension form the series  type $B_k$, with complex form $SO(2k+1,\C)$ and  compact  real form $SO(2k+1,\R)$.\footnote{For the classification simple Lie algebras see any book on Lie algebras/groups;  for the complex case,  e.g., \cite{Knapp:Lie_groups},  thms. 2.84, 2.111 (abstract classification and existence), for a survey of the simple real Lie algebras chap. VI.10. A lucid exposition (in German) gives \cite[501--513]{Brieskorn:LAII}. Weyl had essentially  contributed  to the theory in his book on the {\em Classical Groups} \cite{Weyl:ClassicalGroups}.} 

But even if one accepted such a general structuralist view as a starting point there remained a question for the specification of the space dimension of the group inside the series. 
\begin{quote}
But the number of the dimensions of the world is 4 and not an 
indeterminate $n$. It is a fact that the structure of $\Delta_o$
is quite different for the various dimensionalities $n$. Hence 
the group may serve as a clue by which to discover some cogent reason 
for the dimensionality 4 of the world. What must be brought to 
light, is the distinctive character of one definite group, the 
four-dimensional Lorentz group, either as a group of linear transformations, or as an abstract group. (Hs, p. 22f.)
\end{quote}
The remark that the ``structure of $\Delta_o$
is quite different for the various dimensionalities $n$'' with regard to even or odd complex space dimensions (type $D_k$, resp. $B_k$) strongly qualifies the import of the general structuralist characterization. But   already in the 1920s   Weyl had used  the fact that for the (real) space dimension $n=4$ the universal covering of the unity component of the Lorentz group $SO(1,3)^o$ is the realification of $SL(2,\C)$. The latter  belongs to the first of the $A_k$ series (with complex form $SL(k+1,\C)$). Because of the isomorphism of the initial terms of the series, $A_1 \cong B_1$, this does not imply an exception of Weyl's general statement. We even  may tend to interpret Weyl's otherwise cryptic remark that the structuralist perspective gives a ``at least a partial solution of the problem'' by the observation that the {\em Lorentz group in dimension $n=4$} is, in a rather specific way, the {\em realification of the complex form of one of  the three most elementary non-commutative simple Lie groups} of type $A_1 \cong B_1$.\footnote{The other ``most elementary'' ones belong to the types $C_1, D_2$.} Its compact real form is  $SO(3,\R)$, respectively  the latter's universal cover $SU(2,\C)$.\footnote{See, e.g., \cite[512f.]{Brieskorn:LAII}.}

Weyl stated clearly that the answer cannot be expected by structural considerations alone. The problem is only ``partly one of pure 
mathematics'', the other part is  ``empirical''. But the question itself appeared of utmost importance to him
  \begin{quote}
 We can not claim to have understood 
Nature unless we can establish the uniqueness of the four-dimensional Lorentz  group in this sense. It is a fact that many of the
known laws of nature can at once be generalized	to $n$ dimensions. We  must dig 
deep enough until we hit a layer where this is no longer the case. (Hs, p. 23)
\end{quote}

In 1918 he had given an argument why, in the framework of his new scale gauge geometry, the ``world'' had to be of dimension 4. His  argument had used the construction of the Lagrange density of general relativistic Maxwell theory $\mathcal{L}_{f}= f_{\mu \nu}f^{\mu \nu} \sqrt{|det g|}$, with $f_{\mu \nu}$ the components of curvature of his newly introduced scale/length connection, physically interpreted by him as the electromagnetic field. $\mathcal{L}_{f}$ is scale invariant only in spacetime dimension $n=4$.\footnote{Scale invariance of this Lagrange density presupposes that  the lifting of indices for $f$ compensates the change of  the $\sqrt{|det g|}$ under rescaling. Counting the scale weight of the metric $g_{\mu \nu}$ as 2, scale invariance of $\mathcal{L}_{f}$ holds if and only if  $-4 + \frac{1}{2} 2 n=0 \longleftrightarrow n=4  $,   \cite[p. 31 (p. 37 in the English version)]{Weyl:GuE}. Similarly  for {\em any Yang-Mills Lagrangian}  with Lie algebra valued  2-form $A$, $\mathfrak{L}_{YM}= - \frac{1}{2}tr(A \wedge \ast A)= -\frac{1}{4}A_{\mu_1 \mu_2}A^{\mu_1 \mu_2} \sqrt{|g|}dx^1\wedge \ldots \wedge dx^n$.
\label{fn n=4}}
The shift from scale gauge to phase gauge undermined the importance of this argument. Although it remained correct mathematically, it lost its  convincing power once the scale gauge transformations were relegated from  physics   to the mathematical automorphism group of the theory only. 

Weyl's talk ended with the words:
\begin{quote}
Our question has this in common with most questions of philosophical
nature: it depends on the vague distinction between essential and 
non-essential. Several competing solutions are  thinkable; but it 
may also happen that, once a good solution of the problem is found, it will be of such cogency as to command general recognition.
\end{quote}
This kind of consideration was typical in its openness for his later, mature way of reflecting.

\section{\small Later developments and discussion \label{section discussion}}
In his talk Weyl spanned a huge arc from classical geometry and physics until the 20th century. By surveying this development  he hoped to  be able to identify criteria for the distinction between physical and mathematical aspects in the automorphisms of the respective theories. In the first phase material operations of solid bodies could be used for establishing the elementary basis of congruence. 
  They constituted the fundamental precondition for the possibility of stable trans-subjective measurements   in the time from  Newton  to about Helmholtz (section \ref{section classical}).
 With the shift towards the viewpoint of invariance transformations of physical laws in the last third of the 19th century,  a weak  extension by including orientation reversing isometries became necessary. Although such transformations could no longer be realized by motions of bodies,  an aspect which had been observed in relational form already by Kant, there still remained a chance for an indirect material re-modelling of the additional transformations by mirror transformations and/or point inversions.

On the other hand, mathematical automorphisms of classical geometry were generalized in different versions during the 19th century. Roughly speaking they could be ``any'', depending on the structure mathematicians intended to study. Among these, so Weyl argued, the normalizer of physical automorphisms played a distinguished role.  Normalizer ``in which group''?  Weyl may have thought of  the  general linear transformations or of the differentiable bijective point transformations (diffeomorphisms) of space. In the light of our context, the last interpretation seems more likely.  But one might also consider any normal extension 
 appearing natural or  promising,  by some criterion, as candidate for mathematical automorphisms associated to a given physical  automorphism group.
The reason would be  the same as the one Weyl  gave for similarities with regard to the  congruence relation. A normal extension is ``conservative'' with regard to the relations typical for the   physical automorphisms. 

After the  ``shock of relativity'' the structure of the physical automorphisms became that of a gauge group $\mathcal{G}(P)$ of a principal bundle over spacetime $P \rightarrow M$ with structure group $\Delta_o$  ``read off'' from the then best available theories of matter and its dynamics. For Weyl this was general relativity (gravity and electromagnetism) combined with the Dirac theory of the electron; then $\Delta_o \cong SO(1,3;\R) \times U(1)$ or $SL(2,\C) \times U(1)$.
The invariance transformations of the  physical automorphism group were, in general, no longer representable by material operations, in particular the diffeomorphisms of spacetime. They now took on the form of mathematical transformations of  the theoretical structure designed for representing  natural laws. A part of the transformations, the 
``point dependent rotations'', however, could still  be interpreted  physically in a wider sense,  as  transitions between families of  observer systems. 
In any case, the former clear distinction between mathematical and physical automorphisms  became blurred.  This point was not  discussed by Weyl. Most important for  him was the 
 difference which he could now state between his  1918 scale gauge theory (``only mathematical'')   and  1929 phase gauge theory  (part of ``true'' physics).

In hindsight Weyl's 1948/49 discussion of the physical automorphisms as a gauge group may appear as an anticipation of what was to become a central paradigm for field physics in the last third of the 20th century. 
Gauge theories gained wide recognition with the rise of the  standard model (SM) of elementary particle physics, from the 1970s onward. The  physical automorphisms of the SM is here again a gauge group $\tilde{\mathcal{G}}(P_{SM})$, but now considered over  Minkowski space $\M$ of special relativity, $P_{SM} \longrightarrow \M$. In comparison with Weyl's general relativistic group, the morphisms of the  base are reduced to the Poincar\'e transformations in  $\M$. In this way the 
structure group of the SM is  both, reduced and enlarged, in comparison with the  physical automorphisms of GR considered by Weyl: 
\begin{itemize}
\item[--] The  morphisms of the base $M=\M$ are no longer  the full diffeomorphism group  but are restricted to the linear transformations of the Poincar\'e group. As a  result the 
Lorentz transformations are no longer ``localized'' but are considered as part of the ``global'' transformations  of $\M$.
\item[--] On the other hand and most importantly, the remaining part of Weyl's structure group $U(1)$ is extended to 
 $\tilde{\Delta}_o \cong SU(2)\times U(1)_Y \times SU(3)$  (with the electromagnetic $U(1)$  as residuum after $SU(2)$ symmetry breaking).
\end{itemize}

Moreover, the distinction between physical and mathematical aspects of automorphisms have been  undermined even deeper than in  Weyl's discussion: There remains no 
 chance for a material interpretation of the gauge transformations of the SM, whereas  for Weyl the change of orthonormal frames was still interpretable -- not realizable -- as change between families of ``local'' inertial observers.\footnote{Compare  the discussion in present philosophy of physics on  the ``ontology''  (the physical interpretation) of gauge transformations and the Higgs 
mechanism \cite{Giulini:superselection_rules,Earman:gaugematters,Earman:symmetrybreaking,Lyre:Symmetrien,Healey:Gauging,Rovelli:gauge,Friederich:symmetry,Friederich:2014}.}

There remains the problem of unifying the viewpoints of GR and the standard model of elementary particles in one coherent theoretical framework. One of the  different approaches to achieve this goal envisages a first step towards  an integration  without   a full quantization of gravity:  the strategy to formulate SM fields and their dynamics  on ``curved spaces'' (i.e.  Lorentz manifolds).\footnote{E.g., \cite{Fredenhagen_ea:Where_we_are,Baer/Fredenhagen}.}
Although it is not explicitly discussed by the authors, their work contains   an underlying automorphism group of the theory 
$\mathcal{G}(P)$  with full diffeomorphism group of the base manifold $M$. If one formulates their theory in the orthonormal frame approach, the structure group 
becomes the Lorentz group, or its universal covering, normally extended by the structure group of the SM
\beq G \cong SL(2, \C)\times  SU(2)\times U(1)_Y \times SU(3)  \label{G}
\eeq

This perspective  may even open a new look at Weyl's scale invariance. The Lagrangian of the standard model is basically scale invariant, its fields are formulated  scale covariantly and consistent with their import to scale covariant GR. The only exception is the dimensionful   coefficient $\mu$ of the quadratic term  $\mu\, \Phi \Phi^{\ast}$ of the Higgs field $\Phi$. But it is not difficult to bring the  latent  scale covariance of the SM fields into the open by introducing a second, gravitational and real valued, scalar field $\phi$ of correct scale weight  such that $\mu = \tilde{\mu}\phi^2$ with {\em numerical} coefficient $\tilde{\mu}$. 
With regard to ``global'' scale symmetries over Minkowski space  this is being done, e.g., by \cite{Shapovnikov_ea:2009}. The scale symmetries are ``localized'' in {\em conformal approaches} to SM fields like in 
  and in those working in the framework of {\em  Weyl geometry}\footnote{For the conformal view see, e.g.,  \cite{Meissner/Nicolai,Bars/Steinhardt/Turok:2014}, for the integrable Weyl geometric one  \cite{Nishino/Rajpoot:2011,Nishino/Rajpoot:2007,Quiros:2014,Almeida/Pucheu:2014,Scholz:2015Paving}, for a non-integrable scale connection \cite{Ohanian:2016}.}.
 Inherent in this research is an extension of the automorphism group of the theory to $\tilde{\mathcal{G}}(P)$ like above, but here with structure group  
\beq \tilde{G} \cong R^+ \times  SL(2, \C) \times  SU(2)\times U(1)_Y \times SU(3)  \label{tilde G}
\eeq

But has not such an approach already been refuted by Weyl's argument that quantum physics establishes an ``absolute'' standard of measurement? -- The answer is ``no'', because the gravitational scalar field $\phi$  allows to form scale invariant proportions of quantities which give rise to the measurement values once the unities have been fixed. Moreover, in a particular scale gauge (the one in which the norm of the scalar field is constant) measurement values are expressed  directly, up to a global normalization according to the chosen unit. If one prefers one may describe it, 
 metaphorically, as a kind of ``symmetry breaking'' of scale symmetry by the gravitational scalar field.\footnote{``Metaphorical'' because no phase transition is involved (as far as we can see at the moment) like in the usual understanding of ``spontaneaous symmetry breaking'', cf. \cite{Friederich:gaugesymmetry_breaking}.}

At the moment,  the recent research line based on the  Weyl geometric approach  explores   mathematical models of gravity with or without standard model fields. As long as there are no striking physical consequences  deducible in this framework, which remain unexplained on the basis of  Riemann/Einstein gravity, one has to consider the  group $\tilde{\mathcal{G}}(P)$   of this theory as part of the {\em mathematical automorphisms} in Weyl's language.  But what, if important empirical phenomena are  better explained in this framework than within the underlying Riemann geometric structure with automorphism group $\mathcal{G}(P)$?  -- Then we will be in a situation where the physical automorphisms  of gravity plus SM fields are extended from  $\mathcal{G}(P)$ to $\tilde{\mathcal{G}}(P)$ (with structure group  (\ref{tilde G})) and Weyl's scale extension, the point dependent ``similarities'', become part of the {\em physical automorphism group} again.

This might even give  an  unexpected and intriguing twist to Weyl's  final  question:  Why does the Lorentz group play such a prominent role in the structure group $\Delta_o$, and why $n=4$? As outlined above, his argument of 1918 for the 4-dimensionality of the ``world'', the spacetime manifold of gravity plus electromagnetism,  did not depend on his specific physical interpretation of the scale connection (Weyl's $\varphi_k$) as the potential and the scale curvature (the  $f_{\mu \nu}$) as the components of the Maxwell field. It  is  a {\em structural property of the scale invariance   of the Lagrangian density},  not only for the scale curvature but for all Yang-Mills type Lagrange densities (see fn. \ref{fn n=4}). In particular, the specification of dimension $n=4$ by the scale invariance condition of the Lagrange density works independently of the physical interpretation of the $\varphi_k$.
 If there are reasons to include the (localized) scale transformations in the physical automorphism  group again, Weyl's  argument for $n=4$ will acquire new strength. Together with his observation of the striking simplicity of the Lorentz group as one of the simplest simple Lie groups  we may have the impression that in following Weyl's hints we can come a bit closer to having ``understood Nature'' .


\small
\addcontentsline{toc}{section}{\protect\numberline{}Bibliography}
 \bibliographystyle{apsr}
 \bibliography{a_lit_hist,a_lit_mathsci,a_lit_scholz}

\end{document}